\documentclass[onecolumn,11pt]{revtex4}

\usepackage{graphicx,color} % for figures
\usepackage{amsmath}  
\usepackage{amssymb}
\begin{document}

\title{Short-range correlations and their implications for isospin-dependent modification of nuclear quark distributions}

\author{John Arrington}
\email{johna@anl.gov}
\affiliation{Physics Division, Argonne National Laboratory, Argonne, IL 60439}

\begin{abstract}
The past decade has provided a much clearer picture of the structure of high-momentum
components in nucleons, associated with hard, short-distance interactions between pairs of nucleons.
Recent Jefferson Lab data on light nuclei suggest a connection between these so-called 'short-range correlations'
and the modification of the quark structure of nucleons in the nuclear environment. In light of this
discovery that the detailed nuclear structure is important in describing the nuclear quark distributions,
we examine the potential impact of the isospin-dependent structure of nuclei to see at what level
this might yield flavor-dependent effects in nuclear quark distributions.
\end{abstract}

\smallskip

\maketitle

\section{Introduction}
\label{intro}

A small but important part of the Jefferson Lab (JLab) program involves studies of the high-momentum
structure of nuclei, which requires high-energy interactions to probe cleanly~\cite{arrington12_SRC}.
After the initial observation of identical structure in the high-momentum components of
nuclei at SLAC~\cite{frankfurt93}, measurements at JLab have mapped out the kinematic region where SRCs
dominate~\cite{egiyan03, egiyan06}, mapped out the contribution of SRCs in various light and heavy nuclei
relative to the deuteron~\cite{arrington01, fomin12}, and determined that the SRCs are dominated
(at the $\sim$90\% level) by neutron-proton pairs~\cite{shneor07, subedi08, korover14, hen14}.

Because these high-momentum components are associated with short-distance pairs, and thus high-density
configurations in nuclei, it is natural to think that there could be a connection between SRCs and
the nuclear dependence of nuclear parton distribution functions (pdfs). First it was
noted~\cite{weinstein11} that the common scaling of SRCs and the EMC effect with density might
suggest a connection between the two phenomena. Precise measurements of the EMC effect could be
extended at JLab, taking advantage of the extended scaling in nuclei~\cite{arrington99, arrington06}.
Measurements of both the EMC effect~\cite{seely09} and SRCs~\cite{fomin12} in light nuclei both showed
that $^9$Be deviated from the simple density dependence observed in heavier nuclei. For both effects,
$^9$Be behaved like a rather dense nucleus, even though its average density is quite
low~\cite{carlson14,lu13} due to its significant cluster structure. This led to more detailed
examinations of the correlation, suggesting the possibility that the EMC effect may be related to
either the large momenta or high density of SRCs in nuclei~\cite{arrington12_EMCSRC, hen13, malace14}.

Many calculations of the EMC effect include a binding contribution which explains a significant
portion of the effect~\cite{kulagin04}. The average separation
energy enters into these binding calculations, and a large part of this is associated with SRCs. It has
also been argued that off-shell effects in these high-momentum nucleons could yield additional nucleon
modification~\cite{frankfurt88, gross92, melnitchouk94}. There have been direct measurements using
electron scattering from a deuterium target, using a high-momentum spectator proton to tag scattering
from high-momentum neutrons~\cite{klimenko06}. Existing data show effects consistent with such a
modification of the neutron structure have been observed, but the interpretation is difficult due to
potential final-state interactions. Future measurements are planned at Jefferson
Lab~\cite{e11107_proposal}. More recently, such tagged measurements focusing on
low-momentum nucleons have allowed
for the extraction of the free neutron structure function~\cite{baillie12, tkachenko14}, and have been
used to make the first direct extraction of the nuclear effects in the deuteron relative to the free
proton and neutron~\cite{griffioen15}. At present, these measurements do not have sufficient
precision or kinematic reach to differentiate between different models, but future measurements may
provide further insight into the origin of the EMC effect~\cite{e06113_proposal}.
One can also try to isolate SRCs by
examining scattering at $x>1$, while going to $Q^2$ values where one expects to be sensitive to
the parton distributions in the region dominated by SRCs. Existing data suggest an onset of scaling
at relatively low $Q^2$ for nuclei~\cite{arrington99, arrington06, fomin10}, and an experiment has
been approved to extend such measurement to the maximum $Q^2$ available with the 12 GeV upgrade at
Jefferson Lab~\cite{e06105_proposal}.

Another channel for examining off-shell protons is the comparison of the proton form factors, extracted
by recoil polarization, to the same polarization ratio for proton knockout from a nucleus~\cite{strauch03,
paolone10}. Again, questions about potential final-state interactions have been
raised~\cite{schiavilla05}, but other polarization observables have some sensitivity to these
effects~\cite{malace11} and future measurements will be able to map out the behavior in more
detail~\cite{e11002_proposal}. For both the binding and off-shell contributions, the large
momenta and energies, associated mainly with nucleons in short-range correlations, are the direct
cause of the EMC effect, yielding the observed correlation between the EMC effect and presence of SRCs.

If the EMC effect is driven by the `local density' observed by the struck nucleon, e.g. is due to 
direct quark exchange between nucleons with substantial overlap, then all short-distance
configurations - nn, np, and pp pairs - will contribute. In this case, the presence of these
short-distance configurations is the source of both the EMC effect (due to quark modification in the
overlapping nucleons) and the high-momentum nucleons in SRCs (due to the strong interaction of the
short-range components of the N-N potential. In this case, the EMC-SRC correlation occurs because
both effects have a common origin in these small, dense, two-body configurations
in nuclei. This implies a slightly different prediction for the detailed EMC-SRC correlation, as 
the EMC effect is sensitive to all short-distance NN pairs while SRCs are generated predominantly in
np pairs. In this case, the connection between the EMC effect and SRCs in the nucleus will also depend
on the number of potential nn, np, and pp pairs in the nucleus, implying an A- and Z-dependent
correction to the simple linear correlation assumed in~\cite{weinstein11}. The detailed scaling of the
EMC effect and SRCs were examined in Ref.~\cite{arrington12_EMCSRC}, along with a comparison of the
two pictures above, where the correlation is the result of either local-density or high-virtuality (HV) 
effects. Including a simple pair-counting correction to account for the difference in nn, np, and pp
pairs, the correlation was found to be better (reduced $\chi^2$ value and improved extrapolation to
the deuteron value) under the local density hypothesis. However, the difference is not large and 
the $\chi^2$ test is of somewhat limited use given the potential for correlated errors between the 
measurements on different nuclei, so it is far from a definitive test.

One new direction that has not yet been studied in detail is the flavor-dependence of the EMC effect.
The connection between the EMC effect and presence of SRCs, combined with the isospin structure of
SRCs suggests the potential for significant flavor-dependent effects in nuclear pdfs. This provides
a new observable that can be used to examine the EMC effect and to test models of the nuclear
modification. While this idea has been discussed~\cite{arrington12_SRC, sargsian13}, exisint data on
the unpolarized EMC effect are been insufficient to claim an flavor-dependent effect.

\section{Flavor dependence of the EMC effect}

While the nature of the correlation between the EMC effect and the presence of SRCs is not understood,
most explanations, conventional or exotic, imply that the isospin structure of the correlations would
impact the flavor dependence of the EMC effect. In conventional binding calculations, the fact that the
high-momenta nucleons are generated mainly in n-p pairs implies that a nucleus with significant
neutron excess will have a larger \textit{fraction} of it's protons at high momenta, yielding a larger
separation energy and thus a larger binding effect for protons. The same is true if the correlation
is due to off-shell effects in the high-momentum nucleons of the SRCs. If the EMC effect is driven
by local-density effects, then a similar isospin dependence is natural in nuclei with a large
neutron skin. Nuclei with significant neutron excess typically have a larger neutron radius, meaning
that the lower density surface region will have more neutrons, again yielding a larger average EMC 
effect for protons. Thus, any of these explanations, as well as recent QCD-based models of nuclear
parton distributions~\cite{cloet12_PVEMC}, all predict the same effect: an enhanced EMC effect for
protons (and up-quarks) in neutron rich nuclei.

In the past, the EMC effect was generally assumed to scale with simple bulk properties of the nucleus,
either mass or average nuclear density. The data on light nuclei~\cite{seely09} demonstrate that these
models are not sufficient, and details of the nuclear structure must be accounted for. We examine here
a variety of simple assumptions for the A dependence which account for the nuclear structure. 

The observed EMC-SRC correlation has generally been interpreted based on the idea that the EMC effect
is driven by either local density effects, where overlapping nucleon pairs can have direct quark
exchange or contributions from more exotic states (hidden color or six-quark bags)~\cite{carlson83, sargsian03, miller14},
or by the presence of high-momentum nucleons, due to binding or off-shell effects. These yield
slightly different predictions for the quantitative nature of the relation between the EMC effect
and SRCs, as all NN pairs can contribute to high-density configurations, while the high-momentum
components are dominated by np pairs~\cite{arrington12_SRC}, and will also yield different predictions
for the relative effect on protons and neutrons in the nucleus. All of these yield an increased EMC
effect for protons for nuclei with a neutron excess.

The same enhancement in the minority nucleons occurs in a calculation of nuclear pdfs through the
quark-meson coupling arising in the NJL model~\cite{cloet09_NUTEV, cloet12_PVEMC}. The model results
are obtained for asymmetric nuclear matter,
using a Z/N ratio that matches finite nuclei as a rough estimate for finite nuclei. The nuclear matter
results are likely overestimate the isospin dependence in finite nuclei, although these are mean-field
calculations which do not include realistic nuclear structure for the finite nuclei. So for example,
it does not include any effect connected to short-range correlations or the isospin-dependent density
in a nucleus with a large neutron skin discussed above. Because all of the effects increase the EMC
effect for up-quarks in neutron-rich nuclei, there could be an additional contribution if any of the
nuclear structure effects are also present.

We examine a total of four updated scaling assumptions for the A dependence of the EMC effect:

\begin{itemize}
\item High-momentum fraction: We take the nucleon momentum distribution, $n(k)$, and assume that the
EMC effect scales with the fraction of the distribution above 300~MeV/c. 

\item Average kinetic energy: We take the average kinetic energy of the nucleons, $\langle k^2 \rangle/2M$, evaluated from $n(k)$. 

\item Average density: We evaluate the average density of the nucleus as seen by an individual 
proton or neutron, using the one-body densities for the proton and neutrons

\item Overlap probability: We determine the probability of a proton or neutron being within 1~fm of
another nucleon, based on the two-body densities.

\end{itemize}

The first two quantities, high-momentum fraction and average kinetic energy are related to the idea that
some combination of off-shell effects and the separation energy in conventional binding calculations
are the source of the EMC effect through the presence of high-momentum nucleons. The average density
is the traditional simple model used for scaling of the EMC effect in nuclei. The overlap probability
is one method of examining the impact of local density, using the probability of two nucleons having
significant overlap as an overall measure of the possible contribution to pdf modification due to
the possible contribution from direct quark exchange between the nucleons or more exotic 6-quark or
hidden color configurations. This set of quantities can be calculated from \textit{ab initio} calculations of nuclear structure for light nuclei, providing predictions for the EMC effect in 
light nuclei, which will be measured in an approved experiment at Jefferson Lab~\cite{e12008_proposal}.
We can also examine the difference between the EMC effect for protons and neutrons in asymmetric
nuclei for each of these predictions, which can be examined in the comparison of $^{40}$Ca and
$^{48}$Ca~\cite{e12008_proposal}, as well as in parity-violating electron scattering measurements
from asymmetric nuclei such as $^9$Be and $^{48}$Ca, where the parity-violating asymmetry, related to
the ratio of weak to electromagnetic deep inelastic scattering, is insensitive to any flavor-independent
rescaling of the nuclear pdfs.

\subsection{Results}
\label{sec:results}

For each of the scaling models discussed in the previous section, we calculate the quantity of
interest using the Quantum Monte Carlo calculation~\cite{wiringa14} for all particle-stable nuclei.
The scaling parameter is determined independently for protons and neutrons, and for the unpolarized
case, we combine these with a weight for the number of protons and neutrons and
the relative e-p and e-n cross sections for deep-inelastic scattering at large $x$. The results are
all plotted relative to $^{12}$C, as only the A-dependence and not the overall scale is predicted 
by these simple scaling assumptions.  

\begin{figure}
\centering
%\sidecaption
\includegraphics[angle=90,width=12cm,clip]{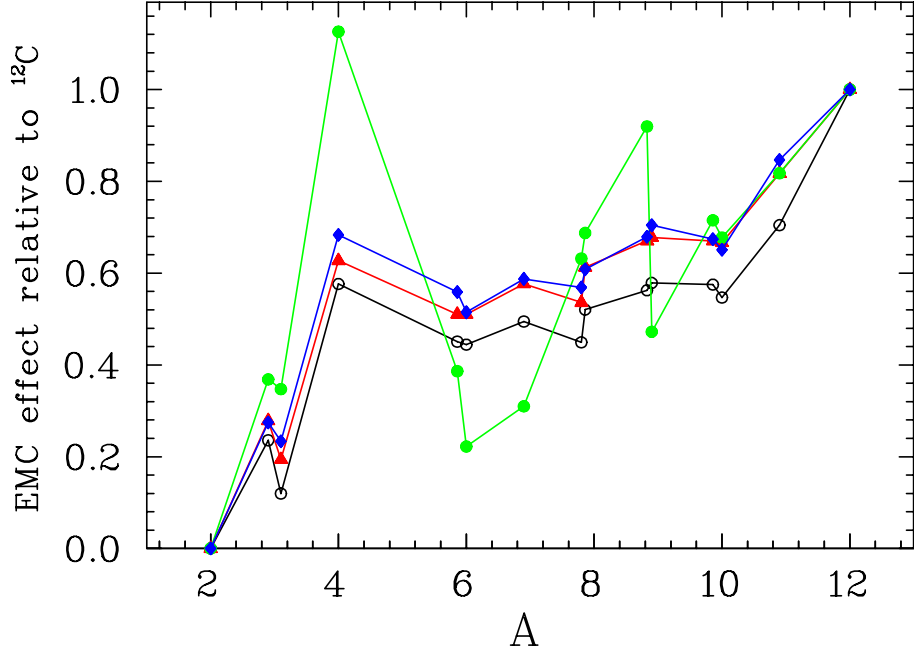}
\caption{A dependence of the EMC effect for the four scaling assumptions: Fraction of momentum
distribution above 300 MeV/c (black open circles), average kinetic energy (red triangles),
average density (green circles), and probability to be within 1~fm of another nucleon (blue diamonds).
The results are shown relative to Carbon, and a slight offset in $A$ is added to differentiate
nuclei with different N/Z ratios.}
\label{fig:EMCunpol}
\end{figure}

Figure~\ref{fig:EMCunpol} shows the prediction for the scaling of the unpolarized EMC effect, taking
the cross section and N/Z weighted average of the proton and neutron predictions. The assumption of
a simple density-dependent scaling is significantly different from the others, and is ruled out by
the existing data on $^3$He, $^4$He, $^9$Be, and $^{12}$C~\cite{seely09}. This is not surprising, as the
simple density dependence picture was already excluded based on these data, but this demonstrates that
accounting for the difference between proton and neutron distributions does not yield a significant
improvement in the prediction.  The other models all show a very similar A dependence for the unpolarized
EMC effect, and are qualitatively consistent with the observation that the modification is large and
similar for $^4$He, $^9$Be, and $^{12}$C. There is a small difference between the predictions, in 
particular for the first scaling model (fraction of high-momentum nucleons), but they are in generally
good qualitative agreement.

\begin{figure}
\centering
%\sidecaption
\includegraphics[angle=90,width=12cm,clip]{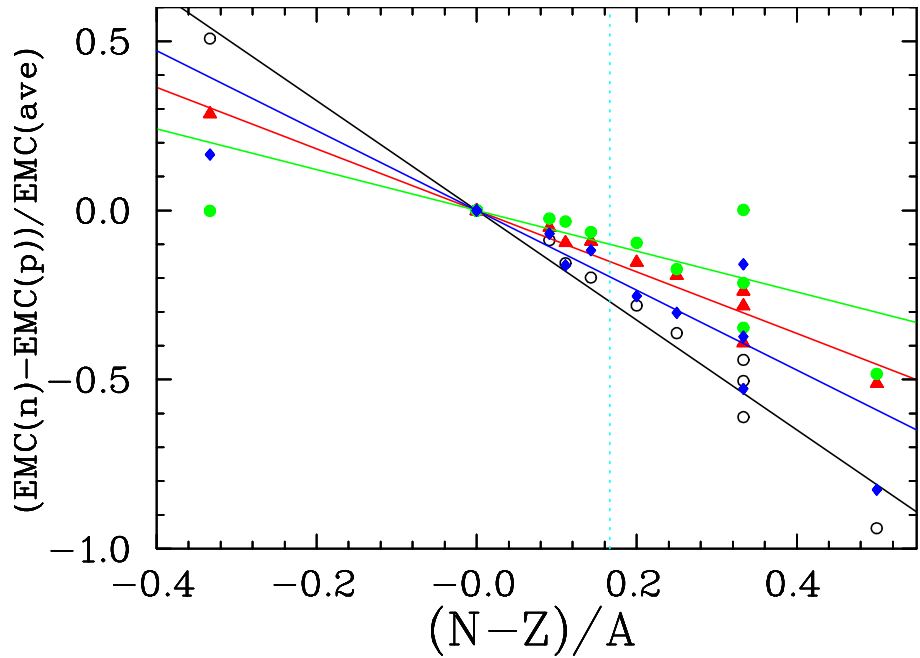}
\caption{Isospin dependence of the EMC effect vs. fractional neutron excess of the nucleus for the
four scaling models (same models as Fig.~\ref{fig:EMCunpol}). The lines are simple unweighted linear
fits. The short-dashed line shows the N/Z value corresponding to $^{48}$Ca.}
\label{fig:EMCisospin}
\end{figure}

Figure~\ref{fig:EMCisospin} shows the fractional difference between the proton and neutron EMC effect
versus the fractional neutron excess of the nuclei. The lines are simple unweighted linear fits which
provide a rough guide as to the overall size of the effects, but note that the behavior is not
linear with fractional neutron excess, in particular for $^3$H and $^3$He. The predicted isospin
dependence shows a significantly larger spread in the predictions for different nuclei, with very large
differences between the EMC effect in protons and neutrons. The largest neutron excess shown in this figure are not easy to access experimentally while $^3$H and $^3$He, which have $|(N-Z)/A|=1/3$, could be measured, but there are technical difficulties using a $^3$H target and the sensitivity will be limited
because the EMC effect in A=3 nuclei is small, making it difficult to extract even a relatively large fractional effect. Measurements on nuclei such as $^9$Be and $^{48}$Ca are straightforward, and these
nuclei have a significant EMC effect as well as the potential to have significant sensitivity to 
different models via their isospin-dependence.

An approved experiment at Jefferson Lab~\cite{e12008_proposal} will measure the unpolarized EMC effect for light nuclei ($^{3,4}$He, $^{6,7}$Li, $^9$Be, $^{10,11}$B, and $^{12}$C) to examine the sensitivity 
of the EMC effect to the detailed nuclear structure in these well-understood nuclei. In addition, it
will have some sensitivity to the isospin dependence through the comparison of light non-isoscalar nuclei
and the comparison of $^{40}$Ca and $^{48}$Ca. However, the unpolarized EMC effect has limited to the
isospin dependence, even where it is large. This is in because there are two effect which partially
cancel in neutron rich nuclei. Increasing the number of neutrons increases the EMC effect for the
protons (up quarks) which dominate the cross section for isoscalar nuclei, but it also increases the
fraction of neutrons (down quarks) which have a decreased EMC effect. Thus, the sensitivity to a change
in N/Z is enhanced when going to proton-rich nuclei and suppressed for neutron-rich nuclei, as seen in the
calculations of Ref.~\cite{cloet09_NUTEV}. 

A much more sensitive test of the isospin dependence of the EMC effect is party-violating deep-inelastic 
electron scattering from non-isoscalar nuclei~\cite{cloet12_PVEMC}. The parity-violating asymmetry is
sensitive to the
interference between Z- and photon-exchange amplitudes relative to the photon-exchange amplitude
squares~\cite{wang14, wang15}, making it sensitive to the ratio of the weak charge to electromagnetic
charge of the quarks. From Ref.~\cite{cloet12_PVEMC}, the $a_2$ parity-violating asymmetry has the
following simplified form when expanding around the $u(x)=d(x)$ limit:
\begin{equation}
a_2(x) = \frac{9}{5} - 4 \sin^2(\theta_W) - \frac{12}{25}\frac{u(x)-d(x)}{u(x)+d(x)}.
\label{eq:1}
\end{equation}
An isoscalar nucleus will have a constant $a_2$ asymmetry, while a non-isoscalar nucleus will
have a small $x$ dependence which is independent of any isospin-independent EMC effect. The asymmetry
expected from any nucleus in the absence of a flavor-dependent EMC effect can therefore be calculated
with minimal uncertainty, based on the small difference between the up- and down-quark distributions
in the excess protons or neutrons. This is relatively well known at low-to-modest values of
$x$~\cite{arrington09, accardi10, accardi11, arrington12_NOVERP, baillie12}, with further measurements
planned to improve our knowledge at larger $x$ values. Any deviation of the measured asymmetry
from the predicted value is sensitive to an isospin-dependent EMC effect that would modify the ratio
of up and down quarks~\cite{cloet12_PVEMC}.

\section{Conclusions}

%Recent measurement of the EMC effect~\cite{seely09} and short-range correlations in light
%nuclei~\cite{fomin12} make it clear that these effects are sensitive to microscopic details of 
%the nuclear structure. One way to probe the microscopic origin is to attempt to try and isolate
%scattering from SRCs, using spectator tagging or the kinematics 

Given what we have learned about the isospin structure of short-range correlations, combined with the
importance of detailed nuclear structure for both SRCs and the EMC effect, it is now difficult to
believe that the EMC effect could be flavor independent, in particular for non-isoscalar nuclei. As
such, a detailed understanding of the EMC effect requires an understanding of the effect in terms
of both A-dependent and isospin-dependent effects.

Measurements of the unpolarized EMC effect in light nuclei (including several non-isoscalar nuclei), 
along with comparisons of $^3$H and $^3$He, along with $^{40}$Ca and $^{48}$Ca will provide some
sensitivity to flavor-dependent effects. The comparison of $^3$H and $^3$He depends on having
reliable understanding of the d/u ratio of the nucleon pdf, and is limited in sensitivity by the
small size of the nuclear effects for A=3. The comparison of the Calcium isotopes should  provide more
sensitivity to isospin-dependent effects, but relies on comparison to a theoretical expectation for
the difference between the EMC effect for A=40 and A=48 in the absence of isospin-dependent effects.
This was assumed to be well understood based on earlier assumptions that the EMC effect had a simple
scaling with the nuclear mass or average density. Because the recent data on light nuclei~\cite{seely09}
demonstrated the impact of more complicated aspects of nuclear structure, including both clustering
and isospin-structure, the comparison may be more sensitive to the assumed scaling in heavy nuclei
than previously assumed.

Pion-induced Drell-Yan scattering has also been proposed as a way to study the flavor dependence of
the EMC effect~\cite{dutta11}. Such measurements would be possible at COMPASS or J-PARC, but there
are no formal plans to make these measurements at present. The cleanest and most precise way to
isolate flavor-dependent effects on the nuclear parton distributions is
through measurements of the parity-violating DIS scattering from nuclei~\cite{cloet12_PVEMC}. The
initial estimates for the isospin dependence of the EMC effect presented here, along with similar
calculations for medium-heavy nuclei, will allow for an optimization of the program, providing a
separation of the A-dependence and isospin-dependence of the EMC effect. This will also provide the
best approach for discriminating between different assumptions about the physics behind the EMC
effect, as illustrated in Fig.~\ref{fig:EMCisospin}.

%\begin{acknowledgement}
This work was supported by the U.S. Department of Energy, Office of Science, Office of Nuclear Physics,
under contract DC-AC02-06CH11357. The author thanks R. Wiringa, I. Cloet, S. Riordan, P. Souder, and
R. Beminiwattha for useful discussions.
%\end{acknowledgement}

\bibliography{FB21_Arrington}

\end{document}